\documentclass[prx, letterpaper, superscriptaddress, amsmath, assume, assume, reprint, floatfix]{revtex4-2}

\usepackage{graphicx}
\usepackage{amsfonts}
\usepackage{dcolumn}
\usepackage{soul}
\usepackage{amsmath}
\usepackage{bm}
\usepackage{xcolor} 
\usepackage[normalem]{ulem}
\usepackage{comment}
\usepackage{braket}

\begin{document}


\title{Radio-frequency electron cascade in semiconductor quantum dots}

\author{Tara Murphy}
    \email{tm763@cam.ac.uk}
    \affiliation{Quantum Motion, 9 Sterling Way, London, N7 9HJ, United Kingdom}
    \affiliation{Cavendish Laboratory, University of Cambridge, J.J. Thomson Avenue, CB3 0HE, United Kingdom}
\author{Lorenzo Peri}
    \affiliation{Quantum Motion, 9 Sterling Way, London, N7 9HJ, United Kingdom}
\author{Giovanni A. Oakes}
    \affiliation{Quantum Motion, 9 Sterling Way, London, N7 9HJ, United Kingdom}
\author{Jacob~F.~Chittock-Wood} 
    \affiliation{Quantum Motion, 9 Sterling Way, London, N7 9HJ, United Kingdom}
    \affiliation{London Centre for Nanotechnology, UCL, London WC1H 0AH, United Kingdom}
\author{Ross~C.~C.~Leon}
    \affiliation{Quantum Motion, 9 Sterling Way, London, N7 9HJ, United Kingdom}
\author{Fabrizio~Berritta}
    \affiliation{Research Laboratory of Electronics, Massachusetts Institute of Technology, Cambridge, MA 02139, USA}
    \affiliation{Institute of Experimental and Applied Physics and Halle-Berlin-Regensburg Cluster of Excellence CCE, University of Regensburg, Regensburg, Germany}
    \altaffiliation{Experimental work performed at the Center for Quantum Devices,
Niels Bohr Institute, University of Copenhagen, Copenhagen, Denmark.}
\author{Amber~Heskes}
    \affiliation{Center for Quantum Devices, Niels Bohr Institute, University of Copenhagen, Copenhagen, Denmark}
\author{David Wise} 
    \affiliation{Quantum Motion, 9 Sterling Way, London, N7 9HJ, United Kingdom}
\author{Anasua~Chatterjee}
    \affiliation{QuTech and Kavli Institute of Nanoscience, Delft University of Technology, Delft, The Netherlands}
\author{Ferdinand Kuemmeth}
    \affiliation{Institute of Experimental and Applied Physics and Halle-Berlin-Regensburg Cluster of Excellence CCE, University of Regensburg, Regensburg, Germany}
    \altaffiliation{Experimental work performed at the Center for Quantum Devices,
Niels Bohr Institute, University of Copenhagen, Copenhagen, Denmark.}
    \affiliation{Q.M Technologies Ltd. (Quantum Machines)}
\author{M. Fernando Gonzalez-Zalba}
    \email{fernando@quantummotion.tech}
    \affiliation{Quantum Motion, 9 Sterling Way, London, N7 9HJ, United Kingdom}
    \affiliation{CIC nanoGUNE Consolider, Tolosa Hiribidea 76, E-20018 Donostia-San Sebastian, Spain} 
    \affiliation{IKERBASQUE, Basque Foundation for Science, E-48011 Bilbao, Spain}

\date{\today}

\begin{abstract}

Electron cotunneling in semiconductor quantum dots is a higher-order tunneling mechanism that connects initial and final charge states through energetically inaccessible virtual intermediate configurations. In quantum-dot arrays, such processes can produce simultaneous charge motion across multiple tunnel junctions and enable transitions that would otherwise be suppressed by Coulomb blockade. Here, building on an earlier proof-of-principle demonstration, we investigate radio-frequency (rf)-driven correlated cotunneling and its application to dispersive quantum-dot readout. Using both theoretical modeling and experimental measurements in two different foundry-fabricated silicon devices, we show that an external rf drive can induce correlated cyclic single-electron transitions, resulting in enhanced charge-polarization of the system. We develop a general theoretical framework for \textit{rf electron-cascade} readout, which exploits cotunneling events in quantum-dot systems with three or more dots to produce a relative signal enhancement, and extend its experimental demonstration to a gate-sensed device architecture. Our results establish the operating principles of \textit{rf electron-cascade} readout and provide a route to increased dispersive readout fidelity in semiconductor quantum computing architectures.

\end{abstract}

\maketitle

\section{Introduction}  

Semiconductor quantum dots (QDs) in the few-charge regime have emerged as a powerful platform for exploring many-body physics, as they enable the realization of artificial fermionic arrays with tunable tunnel couplings~\cite{altman2021quantum, wang2022experimental, kiczynski2022engineering, Jirovec_2026}. For example, QD systems have been shown to exhibit exotic spin and magnetic effects, such as Kondo~\cite{sasaki2000kondo} and Nagaoka~\cite{dehollain2020nagaoka} transitions, and have been used to simulate the Fermi-Hubbard model~\cite{hensgens2017quantum} and the many-body Su-Schrieffer-Heeger model~\cite{kiczynski2022engineering}. Further, simulators that include not only fermionic but also bosonic degrees of freedom, can be constructed by coupling QD arrays to resonators. For example, by representing molecular vibrational modes by microwave resonators, one may simulate molecular physics that include also the nuclear degrees of freedom~\cite{PhysRevB.110.205424}.

QD arrays have also been used for classical logic applications using the computing paradigm known as quantum cellular automata (QCA)~\cite{orlov1997realization}. In QCA, a perturbation of an initial charge state governed by Coulomb repulsion, triggers a series of charge tunneling events that lead to a new ground state corresponding to the solution of a computational problem~\cite{snider1999quantum, porod1999quantum}. In these QD arrays, electron tunneling can occur in a correlated manner known as cotunneling, a higher-order tunneling mechanism in which initial and final charge states are connected through energetically inaccessible virtual intermediate configurations \cite{Gustavsson2008Virtual,Tran2005MultipleCotunneling}. 
Such cotunneling events are well established in semiconductor QD systems~\cite{de2001electron, golovach2004transport, zumbuhl2004cotunneling}, and have been leveraged to extend the measurement range of a charge sensor~\cite{van2021electron} which finds applications in quantum computing with spin qubits~\cite{Xue2022, Noiri2022, Mills2022, Abraham2026, Undseth2026}. So far these cotunneling effects and applications have been studied using direct current (dc) transport measurements, but their applications in the high frequency domain are still lacking. 

Here, we leverage electron cotunneling at high frequencies to describe a new method to readout the charge and spin state of semiconductor quantum dots, \textit{the radio-frequency electron cascade}. The cascade consists of correlated electron-tunnelling events in which a charge transition within a target DQD induces an additional transition in a nearby sensor dot, thereby enhancing the measured single-electron alternating-current (ac) signal. In this work, building on the original demonstration of rf electron-cascade readout~\cite{Chittock-Wood2026}, we investigate the physical mechanism underlying the effect and develop a theoretical framework describing the resulting signal amplification.  We then demonstrate the mechanism in two different foundry-fabricated silicon devices, including a second architecture in which the sensor is probed through a gate rather than through its reservoir. The measurements yield lower-bound amplitude signal-to-noise ratio (SNR) enhancements of $\sqrt{A}\approx 50$ and $\sqrt{A}\approx 40$ in the two implementations, and are supported by charge-stability simulations consistent with the proposed cascade mechanism. Finally, we discuss how this mechanism could be used to enhance dispersive readout in future semiconductor devices, including Majorana~\cite{Aghaee2025} and germanium QD platforms~\cite{dijkema2026}.

\section{RF ELECTRON CASCADE: CONCEPT}

As QD technology matures, new applications of QD systems emerge, particularly in the high-frequency domain. Here, elastic charge transitions between QDs are exploited in low-power cryogenic electronic devices, such as microwave detectors~\cite{oppliger2026}, masers~\cite{day_2024_masers},
parametric amplifiers~\cite{Cochrane2022a, elhomsy2023broadband}, thermometers~\cite{Ahmed2018b, blanchet2022rfcbt}, frequency mixers~\cite{hogg2023semiconductor} and multipliers~\cite{oakes2023quantum}.
 
\begin{figure}
    \centering
    \includegraphics[width=\linewidth]{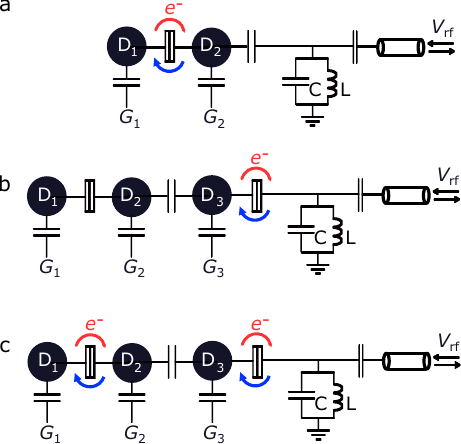}
    \caption{Schematic of dispersive reflectometry techniques to probe the charge state of a DQD. (a) \textit{In-situ} dispersive readout measures changes in the polarizability of a DQD system as seen from a gate electrode as it exchanges electrons between $D_1$ and $D_2$. (b) Dispersive charge sensing of a nearby DQD is done by measuring changes in the polarizability of a SEB as a consequence of charge state changes in a nearby DQD. In the diagram, the SEB has high polarizability, i.e. the charge tunneling between the SEB and the reservoir occurs. A change in charge state of the DQD shifts the SEB operating point reducing or even quenching the ac single-electron current. We note that SEBs can also be probed via a gate electrode ($G_3$) where the detected charge is the continuous charge on that gate. (c) Dispersive \textit{rf cascade}, where an ICT in $D_1$--$D_2$ triggers a correlated transition between $D_3$ and the reservoir. This amplifies the ac current collected at resonator, enhancing the effective polarizability of the $D_1$--$D_2$ transition by combining dispersive charge sensing and \textit{in-situ} dispersive readout. We note that cascade can also be probed via a gate electrode ($G_3$) where the detected charge is the continuous charge on that gate.}
    \label{fig:Fig1}
\end{figure}

Among high-frequency applications, alternating current (ac) measurements using dispersive rf reflectometry readout have emerged as a promising, low-power, and scalable technique to measure the charge and spin state of QDs~\cite{Vigneau_2023}. In its most compact and scalable implementation, often referred to as \textit{in-situ} dispersive readout, an ac voltage is applied to a qubit gate to probe the ac polarizability of a double QD system, see Fig.~\ref{fig:Fig1}(a). Changes in this polarizability result in changes in the gate current which can be used to infer the quantum state of the system~\cite{Petersson2010, urdampilleta2019gate, zheng2019rapid, crippa2019gate}. However, despite being the most compact spin qubit readout method, \textit{in-situ} readout suffers from a comparable lower signal compared to other sensing methods, partially due to the relatively small gate current generated by the event to be sensed~\cite{West2019, prabowo2024intrinsicsnr}.
To address this challenge, dispersive charge sensing, that utilises the polarizability of a dispersive charge sensor such as the single-electron box (SEB), can be utilized. In the case of the SEB, the sensing current is constituted by cyclic single-electron dot-reservoir transitions (DRTs) at the rf excitation frequency; a current that can be larger than the gate current in \textit{in-situ} implementations. However, dispersive charge sensing does not measure the ac polarizability of the target system and requires this separate charge sensor and hence a larger device footprint to enable charge and spin state readout of a target system~\cite{chanrion2020charge, borjans2021spin, Oakes2023, hogg2023single}, as illustrated in Fig.~\ref{fig:Fig1}(b).

A third approach, dubbed the \textit{rf electron-cascade readout}, see Fig.~\ref{fig:Fig1}(c), enables measuring the ac polarizability of the target DQD overcoming the low signal limitations of \textit{in-situ} dispersive readout. The sensing architecture consists of a DQD that is strongly electrostatically coupled to an additional QD which additionally is tunnel-coupled to a charge reservoir. Under appropriate gate-voltage conditions, an interdot charge transition (ICT) within the DQD shifts the electrochemical potential of the additional dot across the Fermi level of the reservoir, thereby triggering a correlated dot--reservoir transition (DRT), as illustrated in Fig.~\ref{fig:Fig1}(c). This charge rearrangement constitutes an \textit{rf electron cascade}. By linking the occupation of the additional dot to that of the DQD, the cascade converts the comparatively weak dispersive response of an ICT into the larger ac response associated with a DRT. The mechanism therefore combines the strong signal available from a dispersive charge sensor with the direct sensitivity to a DQD transition characteristic of \textit{in-situ} readout. In the following, we derive the conditions required for cascade amplification and determine its dependence on drive amplitude, temperature, tunnel coupling, and capacitive lever arms.

\section{RF ELECTRON CASCADE: THEORY}
\label{sec:theory}
We define an \textit{rf electron cascade} as an rf-driven charge rearrangement process in which an ICT in a target DQD electrostatically triggers a correlated charge transition in nearby coupled QDs. In this regime, the charge motion of the adjacent dot becomes correlated with the charge occupation of the DQD of interest, so that a transition in the target system is converted into a larger effective cross capacitive shift of radio-frequency sensor dots in the system. The result is an enhanced dispersive rf response compared with standard \textit{in-situ} readout, where only the polarizability of the target ICT is probed directly.

To understand the \textit{rf electron cascade} and why it results in an enhanced rf response with respect to standard \textit{in-situ} readout, we compare the two configurations illustrated in Fig.~\ref{fig:Fig1}(a) and (c).

\paragraph{In-situ regime.}
We first consider a DQD consisting of two dots $D_i$,
$i=1,2$, each capacitively coupled to a gate $G_i$, with $D_2$ further coupled to the sensing resonator. 
Varying the DQD detuning $\varepsilon \propto V_{G_1} - V_{G_2}$, with $V_{G_i}$ being the voltage applied on gate $G_i$, makes one site more energetically favorable, resulting in an ICT between $D_1$ and $D_2$, with the ground state evolving from $(0,1)$ to $(1,0)$.
The rf excitation, probes the ac polarizability associated with this charge rearrangement directly.

\paragraph{rf electron cascade.}
We now extend the system to a triple QD configuration with dots $D_i$
($i=1,2,3$) controlled by gates $G_i$. We use this minimal three-dot system throughout the remainder of the theory section to illustrate the cascade mechanism, although the same principle can arise in other QD architectures, as demonstrated in Sec.~\ref{sec:exp2}, where a four QD device is used. 
In the three dot setup, $D_3$ is capacitively coupled to the resonator, and a finite tunnel rate between the reservoir and $D_3$ allows particles to exchange between the two. Thus, $D_3$ acts effectively as an SEB, while $D_1$ and $D_2$ form a DQD to be sensed. We note that the concept also works if $D_3$ is probed by the resonator via $G_3$ where in this case the detected charge is the continuous charge on that gate rather than primarily the discrete single-electron ac current collected at the reservoir. 

We consider the configuration in which $D_1$ and $D_3$ are occupied, corresponding to the charge state $(1,0,N)$, and vary the DQD detuning $\varepsilon \propto V_{G_1}-V_{G_2}$. As the detuning is swept, the relative energies of $D_1$ and $D_2$ change, driving an ICT between the two dots. Owing to the strong mutual capacitance between $D_2$ and $D_3$, occupation of $D_2$ simultaneously raises the electrochemical potential of $D_3$. When this shift is sufficiently large to move the $D_3$ level across the reservoir Fermi level, the $D_1$--$D_2$ ICT is accompanied by electron exchange between $D_3$ and the resonator reservoir. The system therefore undergoes the correlated cyclic transition $(1,0,N)\leftrightarrow(0,1,N-1)$ driven by the rf signal.

This rearrangement proceeds via two simultaneous tunnelling events, between $D_1$ and $D_2$, and between $D_3$ and the reservoir, without occupying the $(0,1,N)$ or $(1,0,N-1)$ configurations. Strong interdot electrostatic coupling makes both of these states energetically inaccessible, so the transition is a correlated cotunnelling process~\cite{Gustavsson2008Virtual,Tran2005MultipleCotunneling},and shown in the eigenvalue spectra of the Hubbard Hamiltonian in Fig.~\ref{fig:Figure1}(a)--(b) (see Appendix~\ref{app:lorenzo} and Eq.~(\ref{eq:H_hub})).

In the \textit{rf electron-cascade} regime, this correlated charge rearrangement converts a single ICT into a larger effective charge displacement, thereby enhancing the detected rf response. In particular, the occupation of $D_3$, which is directly tunnel-coupled to the reservoir, becomes correlated with the charge state of the remote $D_1$--$D_2$ system. Consequently, as $\varepsilon$ is driven through the ICT, $D_3$ simultaneously undergoes a DRT with the reservoir, producing the enhanced dispersive signal. The resulting SNR amplification is visible in the simulated charge stability diagrams in Fig.~\ref{fig:Figure1}(c)-(d) for the \text{in-situ} and \text{rf electron cascade}, respectively, and the ICT signal amplitude in Fig.~\ref{fig:Figure1}(e). 
\begin{figure}[h!]
\includegraphics[width=\linewidth]{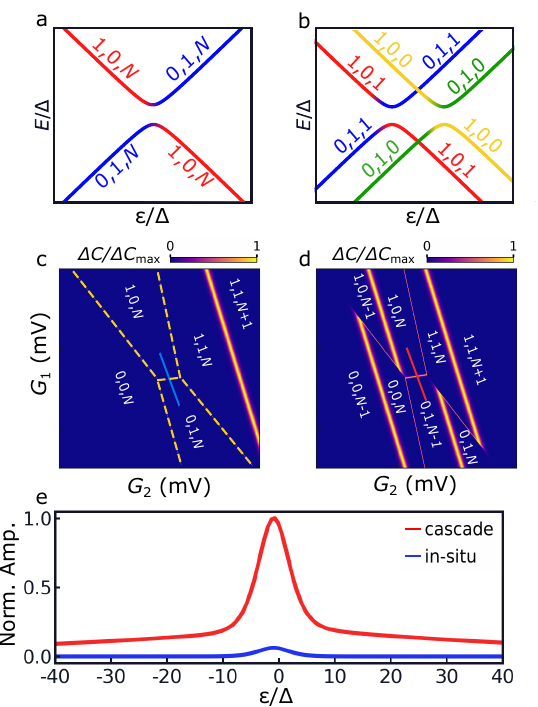}
\caption{
(a), (b) Eigenvalue spectra of the QD system as a function of the detuning $\varepsilon$, normalized by the full avoided-crossing gap $\Delta$ for the \textit{in-situ} configuration (a) and the \textit{rf electron cascade} configuration (b). Charge configurations are indicated for the relevant eigenstates. (c), (d) Corresponding normalized charge stability diagrams showing the ICT without (c) and with (d) the cascade process. Transitions have been highlighted by the yellow dashed lines. The detuning axis has been highlighted by a red and blue line on both graphs. (e) Normalized rf response as a function of reduced detuning, comparing the \textit{rf electron cascade} (red) and \textit{in-situ} (blue) configurations, demonstrating the enhanced dispersive signal produced by the \textit{rf electron cascade}. Simulation parameters are provided in Appendix~\ref{app:C_matrix}.}
\label{fig:Figure1}
\end{figure}

The change in quantum capacitance measured is determined by changes in the average occupations of the dots, weighted by their respective lever arms
$\alpha_{Ri}$ between the reservoir and dot $D_i$,
\begin{equation}
dq = e \sum_{i=1}^{3} \alpha_{Ri}\, d\langle n_i \rangle ,
\label{eq:dq_full}
\end{equation}
where $\langle n_i \rangle$ denotes the expectation value of the occupation of dot $D_i$ at equilibrium. We consider the regime in which the rf drive frequency is significantly lower than the characteristic tunnel rates of the system. Therefore, we operate in the adiabatic regime where the charge distribution continuously relaxes to its instantaneous equilibrium state.

Across the \textit{rf electron cascade} transition $[(0,1,N-1)\leftrightarrow(1,0,N)]$, for a small change in charge occupation probability, $dP$,  the occupations change in a correlated manner, such that
$d\langle n_1\rangle = -dP$,
$d\langle n_2\rangle = +dP$, and
$d\langle n_3\rangle = -dP$. Substituting into Eq.~(\ref{eq:dq_full}) gives
\begin{equation}
dq = e\,\alpha_{\rm casc}\, dP,
\label{eq:dP_main}
\end{equation}
with the effective cascade lever arm
\begin{equation}
\alpha_{\rm casc}
= 1 -\alpha_{R3} + \alpha_{R2} - \alpha_{R1},
\label{eq:alpha_casc}
\end{equation}
with $1 - \alpha_{R3}$ being the lever arm between the cascaded dot and resonator reservoir, $\alpha_{R2}$ and $\alpha_{R1}$ being the lever arm between the resonator reservoir and $D_2$ and $D_1$, respectively. Therefore, the capacitance arising from the cascade may be modeled as an effective DRT of $D_3$ with renormalised lever arm $\alpha_{\rm casc}$~\cite{Peri_2024}. Further details of the equivalent lever arm extraction are found in Appendix~\ref{app:lorenzo}.

The measured reflected rf voltage is directly proportional to the
quantum capacitance of the reservoir-coupled dot~\cite{Peri_2024_ul}.
For an rf excitation
\[
v(t)=V_0+V_{\rm dev}\sin(\omega t).
\]

The effective capacitance associated with a transition \(j\) is then obtained
from the in-phase Fourier component of the charge response,
\begin{equation}
C_j =
\frac{e\alpha_j \omega}{\pi V_{\rm dev}}
\int_0^{2\pi/\omega}
n_j\!\left(v(\omega t)\right)\sin\omega t\,dt ,
\label{eq:Cq_general}
\end{equation}
where \(n_j\) is the dimensionless occupation associated with the transition, which can take different functional forms depending on the type of transition~\cite{Maman2020},
and \(\alpha_j\) is the corresponding effective lever arm and $V_{\rm dev}$ is the drive amplitude.

In the case of the cascade, the transition can be modeled as an effective DRT with lever arm \(\alpha_{\rm casc}\) (see Appendix~\ref{app:lorenzo}). The corresponding occupation follows a Fermi--Dirac distribution,
\begin{equation}
n_{\rm casc}(v)
=
\frac{1}{
\exp\!\left[
\dfrac{\alpha_{\rm casc} e v}{k_{\rm B}T}
\right]
+1
}
\label{eq:nR_DRT}
\end{equation}

\noindent where $k_BT$ is the thermal energy.
For the \textit{in-situ}, the occupation is set by the two-level avoided crossing \cite{Maman2020},
\begin{equation}
n_{\rm in\text{-}situ}(v)
=
\frac{1}{2}
\left[
1+
\frac{\varepsilon_{\rm in\text{-}situ}(v)}{E(v)}
\tanh\!\left(
\frac{E(v)}{2k_{\rm B}T}
\right)
\right],
\label{eq:nR_ICT}
\end{equation}

with $\varepsilon_{\rm in\text{-}situ}(t) = \alpha_{\rm in\text{-}situ}\,e\,v(t)$,
$E(t) = \sqrt{\varepsilon_{\rm in\text{-}situ}(t)^2+\Delta^2}$, and $\Delta$ is the minimum energy splitting at the avoided
crossing, corresponding to an interdot hopping matrix element
$t_c=\Delta/2$.
The cascade and in-situ capacitances are obtained by substituting
Eqs.~(\ref{eq:nR_DRT}) and (\ref{eq:nR_ICT}) into
Eq.~(\ref{eq:Cq_general}) with
\(\alpha_j=\alpha_{\rm casc}\) and
\(\alpha_j=\alpha_{\rm in\text{-}situ}\), respectively.

The cascade amplification factor is defined as the ratio of the change in
reflected rf power in the cascade configuration to that in the
\textit{in-situ} configuration. Since the change in reflected rf power is
proportional to the square of the capacitance change~\cite{Vigneau_2023}, this
gives
\begin{equation}
A =
\left(
\frac{C_{\rm casc}}{C_{\rm in\text{-}situ}}
\right)^2 
\label{eq:A_ratio_of_projections}
\end{equation}
where $C_{\rm casc}$ and $C_{\rm in\text{-}situ}$ are calculated by
substituting Eqs.~(\ref{eq:nR_DRT}) and (\ref{eq:nR_ICT}), respectively, into
Eq.~(\ref{eq:Cq_general}). Here, $A$ quantifies the relative power-response gain of cascade readout with
respect to direct \textit{in-situ} sensing, rather than the absolute
sensitivity of the cascade transition. In particular, an increase in $A$ may
result either from an increase in $C_{\rm casc}$ or from a reduction in
$C_{\rm in\text{-}situ}$. Provided that the resonator
transduction and measurement noise remain unchanged between the two
configurations, this quantity also corresponds to the expected amplitude-SNR
enhancement between the cascade and \textit{in-situ} measurement modes. The corresponding amplitude-response gain is $\sqrt{A}=|C_{\rm casc}/C_{\rm in\text{-}situ}|$.

\begin{figure}
    \centering
    \includegraphics[width=\linewidth]{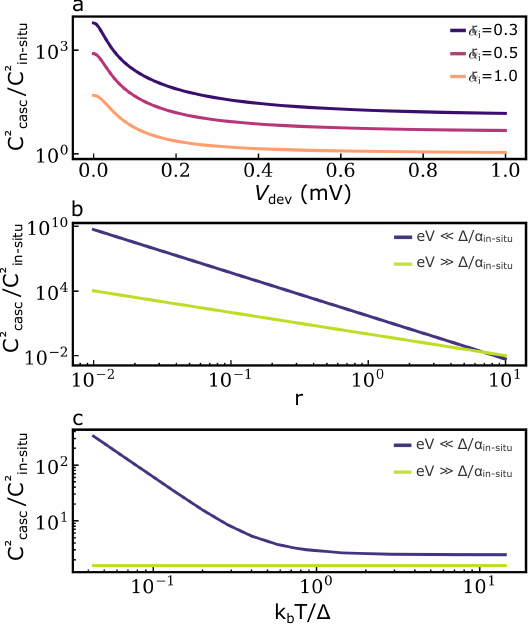}
    \caption{
    Numerically calculated cascade power-response gain
    $A=(C_{\rm casc}/C_{\rm in\text{-}situ})^2$.
    (a) Dependence of $A$ on the device-level rf amplitude
    $V_{\rm dev}$, evaluated at fixed $T=50~\mathrm{mK}$,
    $\Delta=60~\mu\mathrm{eV}$, and $\alpha_R=0.10$, for
    different values of the lever-arm ratio
    $r=\alpha_{\rm in\text{-}situ}/\alpha_{\rm casc}$.
    (b) Dependence of $A$ on $r$. In the linear-response regime,
    defined by
    $\alpha_{\rm casc}eV_{\rm dev}\ll k_{\rm B}T$ and
    $\alpha_{\rm in\text{-}situ}eV_{\rm dev}\ll\Delta$,
    the gain scales as $r^{-4}$. In the large-drive regime,
    defined by
    $\alpha_{\rm casc}eV_{\rm dev}\gg k_{\rm B}T$ and
    $\alpha_{\rm in\text{-}situ}eV_{\rm dev}\gg\Delta$,
    the gain approaches the $r^{-2}$ limit.
    (c) Dependence of $A$ on $k_{\rm B}T/\Delta$. In the
    linear-response regime, the gain decreases with increasing
    temperature and saturates for $k_{\rm B}T\gg\Delta$; in the
    large-drive regime, it is approximately independent of
    temperature. The complete set of numerical parameters is given
    in Appendix~B.
    }
    \label{fig:Fig3}
\end{figure}

We assume that the resonator parameters, including the internal quality
factor, matching coefficient, and resonator capacitance, remain the same for
both sensing implementations. The ratio
$r\equiv\alpha_{\rm in\text{-}situ}/\alpha_{\rm casc}$ parametrizes the
difference in effective lever arm between the two modes of operation. The
conditions under which $A>1$ therefore depend on both $r$ and
$\Delta/k_{\rm B}T$, as explored in Fig.~\ref{fig:Fig3} and
Appendix~\ref{app:limits}.

Figure~\ref{fig:Fig3} shows $A$ numerically evaluated from 
Eq.~(\ref{eq:A_ratio_of_projections}) as a function of the key experimental
parameters; full analytic derivations are given in Appendix~\ref{app:limits}. We distinguish two limiting drive regimes. The linear-response
regime is defined by
\begin{equation}
    \alpha_{\rm casc} e V_{\rm dev} \ll k_{\rm B}T,
    \qquad
    \alpha_{\rm in\text{-}situ} e V_{\rm dev} \ll \Delta,
    \label{eq:linear_drive_conditions}
\end{equation}
such that the rf excitation is small compared with the intrinsic
width of both DRT and ICT transitions. Conversely, the large-drive regime is
defined by
\begin{equation}
    \alpha_{\rm casc} e V_{\rm dev} \gg k_{\rm B}T,
    \qquad
    \alpha_{\rm in\text{-}situ} e V_{\rm dev} \gg \Delta,
    \label{eq:large_drive_conditions}
\end{equation}
such that the rf excitation produces a full transition within each rf cycle.

Panel (a) shows the dependence on the drive amplitude $V_{\rm dev}$ for different values of the ratio $r$. For each value the amplification is maximised at small drive and decreases as
$V_{\rm dev}$ increases and eventually saturates. Panel (b) shows the dependence on the lever-arm ratio $r$. In the linear-response regime defined by
Eq.~(\ref{eq:linear_drive_conditions}), the capacitances are
proportional to the squares of their respective effective lever
arms, yielding,
\begin{equation}
A^{\rm lin}
= \frac{1}{r^4}
\left(\frac{\Delta}{2k_\text{B}T}\right)^2 \tanh^{-2}(\Delta/2k_\text{B}T),
\label{eq:A_lin}
\end{equation}
giving the steep $r^{-4}$ dependence visible as the blue curve. In the large-drive regime defined by
Eq.~(\ref{eq:large_drive_conditions}), both occupation functions
may be approximated as step functions~\cite{Maman2020}. The corresponding
capacitances are then proportional to their respective effective
lever arms, yielding
\begin{equation}
A^{\rm large} = \frac{1}{r^{2}},
\label{eq:A_large}
\end{equation}
shown as the green curve. Panel~(c) shows the dependence on $k_{\rm B}T/\Delta$. For $\Delta\gg k_{\rm B}T$, the linear-response
power-response gain grows as
$(\Delta/k_{\rm B}T)^2$. Importantly, this increase does not
correspond to an enhancement of the absolute cascade response. In the
linear-response regime,
\begin{equation}
C_{\rm casc}^{\rm lin}
=
\frac{\alpha_{\rm casc}^2 e^2}{4k_{\rm B}T},
\end{equation}
which is independent of $\Delta$, whereas for $\Delta \gg k_{\rm B}T$,
\begin{equation}
C_{\rm in\text{-}situ}^{\rm lin}
\approx
\frac{\alpha_{\rm in\text{-}situ}^2 e^2}{2\Delta}.
\end{equation}

The corresponding derivations are provided in App.~\ref{app:Cq_integral}. The increase in $A$ with tunnel coupling therefore arises because increasing
$\Delta$ suppresses the direct ICT response, increasing the relative advantage
of cascade readout. By contrast, reducing the temperature, increases
$C_{\rm casc}$ and therefore improves both the cascade response
and its gain relative to direct \textit{in-situ} sensing. In the large-drive regime, shown by the green curve, the amplification is largely insensitive to $\Delta/k_{\rm B}T$ because
both the in-situ and cascade occupations are driven over voltage
ranges broad compared with their respective intrinsic transition
widths. The gain therefore approaches
$A^{\rm large}=r^{-2}$.

Together, this analysis shows that the electron cascade effectively
maps an ICT onto a DRT, resulting in an amplified dispersive response.
The derived scaling relations identify the key parameters governing this
amplification, namely the lever-arm ratio, temperature, and tunnel coupling,
and provide clear experimental signatures of the cascade regime. We now turn to experiment to demonstrate the cascade mechanism and the resulting signal enhancement in two different semiconductor QD architectures.

\section{Experimental Results}
\label{sec:exp}
\subsection{Device A: Planar silicon (SiMOS)}
\label{sec:exp1}

We first revisit the original reservoir-sensed demonstration of rf electron-cascade readout reported in Ref.~\cite{Chittock-Wood2026}. Here, we use these measurements as a benchmark for the theoretical framework developed in Sec.~\ref{sec:theory} and compare the observed charge configurations and signal enhancement with corresponding simulations.

\begin{figure}[h!]
\includegraphics[width=\linewidth]{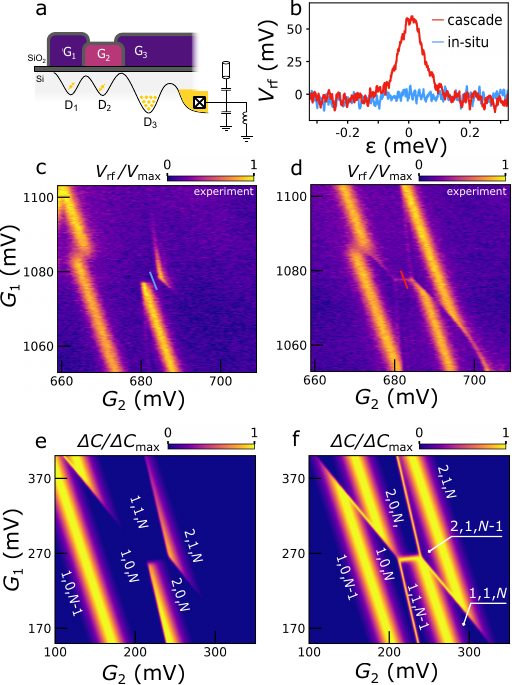}
\caption{
(a) Illustration of the SiMOS device, comprising a DQD
$D_1$--$D_2$ adjacent to a sensor dot $D_3$.
The reservoir of $D_3$ is wire-bonded to an off-chip
superconducting spiral resonator for reflectometry readout. (b) Demodulated reflectometry voltage $V_{\rm rf}$ measured along the DQD
detuning axis with the cascade enabled (red) and suppressed (blue). The two
traces were acquired at the same gate-bias conditions; for the blue trace, a
finite magnetic field suppresses the cascade through Pauli spin blockade.
The comparison yields a lower-bound amplitude-SNR enhancement of
$\sqrt{A}\approx 50$. (c), (d) Experimental charge-stability diagrams measured for configurations
in which the sensor dot is misaligned (c) or aligned (d) with the DQD ICT.
When aligned, the ICT triggers a DRT in the sensor,
producing the enhanced signal associated with the cascade. (e), (f) Simulated charge-stability diagrams corresponding to the
experimental configurations in (c) and (d), respectively. The sensor-dot
voltage is set to $V_{G_3}=320$~mV in (e) and $V_{G_3}=300$~mV in (f). The simulation
parameters are listed in Appendix~\ref{app:C_matrix}. The experimental data
in panels (c) and (d) are reproduced from Ref.~\cite{Chittock-Wood2026}.
}
\label{fig:Figure3}
\end{figure}

To experimentally demonstrate the electron cascade process, we investigate two different voltage configurations in a device fabricated at IMEC using an industry-compatible 300~mm silicon metal-oxide-semiconductor (SiMOS) process with overlapping gate technology \cite{Stuyck21}. Three layers of polycrystalline-silicon gates define the array of QDs, and their electrostatic potentials are controlled via gate voltages $G_i$. A simple cross section of the device is given in Fig.~\ref{fig:Figure3}(a).

We operate the device as a two-electron DQD, consisting of two coupled QDs, which we define as $D_1$--$D_2$, and an adjacent multi-electron sensor dot ($D_3$). An ohmic contact to the reservoir is connected to an off-chip superconducting NbTi spiral inductor, forming an LC tank circuit with a resonance frequency of 512~MHz. An rf tone is applied to this ohmic contact, and the reflected signal is demodulated for readout. Further details of the experimental setup can be found in Ref.~\cite{Chittock-Wood2026}.

For the experimental line cuts in Fig.~\ref{fig:Figure3}(b), the signal-to-noise ratio was extracted from the demodulated rf voltage trace as $\mathrm{SNR}_{\rm casc}= \left(V_{\rm peak}-V_{\rm bg}\right)^2/\sigma_{\rm bg}^2$
where $V_{\rm peak}$ is the cascade line-cut point with the largest deviation from the background level, $V_{\rm bg}$ is the background/reference voltage level, and $\sigma_{\rm bg}$ is the standard deviation of the corresponding background line cut without the cascade process. The quoted amplification factor is then calculated relative to the \textit{in-situ} reference value, $\mathrm{SNR}_{in-situ}=0.5$, as $A=\mathrm{SNR}_{\rm casc}/\mathrm{SNR}_{in-situ}$


Figures~\ref{fig:Figure3}(c)-(d) show charge stability diagrams for two voltage configurations that either suppress (c) or enable (d) the electron cascade event along the ICT of the DQD. By altering the bias on $G_3$,  which primarily controls the gate-induced charge of the sensor dot $D_3$, we can tune the system in and out of the \textit{rf electron cascade}. 

Consequently, the ICT signal is significantly enhanced when the sensor is aligned with the cascade transition, Fig.~\ref{fig:Figure3}(d), while it is reduced to the noise level in the misaligned configuration shown in Fig.~\ref{fig:Figure3}(c). Separately, Fig.~\ref{fig:Figure3}(b) compares high-resolution line cuts acquired at the same gate-bias conditions, with the cascade enabled in the red trace and suppressed in the blue trace by applying a finite magnetic field and inducing Pauli spin blockade. This comparison yields a lower-bound amplitude-SNR enhancement of $\sqrt{A}\approx50$.

To confirm that an electron cascade event underpins this enhancement in dispersive signal, we simulate the rf response of the device and determine the charge configurations throughout the CSD (see Appendix~\ref{app:C_matrix} for details). These simulations are given in Figures~\ref{fig:Figure3}(e)-(f), together with the relevant charge occupations under each voltage configuration. In the cascade configuration, Fig.~\ref{fig:Figure3}(d) and (f), the ground state transition proceeds from $(0, 2,N-1)$ to $(1,1,N)$, corresponding to the amplified signal. Here, N denotes the number of electrons in the multi-electron sensor dot.

By contrast, Fig.~\ref{fig:Figure3}(c) and (e) represent a voltage configuration in which $G_{3}$ is tuned to bias $D_3$ deep in the Coulomb blockade, such that the electrostatic shift arising from the $D_1$--$D_2$ ICT is not enough to cause a reservoir transition.
Consequently, in this configuration, the ground state changes from $(0,1,N)$ to $(1,0,N)$, which does not trigger a cascade. These observations align with our theoretical model and corroborate the cascade-driven amplification mechanism.

\subsection{Device B: Fully-depleted silicon-on-insulator etched nanowire}
\label{sec:exp2}

\begin{figure}[h!]
    \centering
    \includegraphics[width=\linewidth]{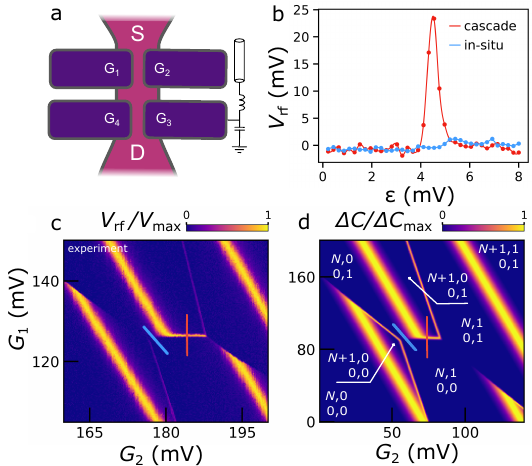}
    \caption{
(a)  Illustration of the FDSOI 2$\times$2 quantum-dot array. Below gates $G_1$, $G_2$, and $G_3$, a triple dot is activated by control voltages. 
The sensor dot is read out by rf reflectometry using a carrier tone applied through an off-chip SMD inductor wire-bonded to gate $G_3$.
(b) Demodulated rf reflectometry voltage signal $V_{\rm rf}$ in red (blue) along the linecut from (c) with (without) the cascade process.
(c) Measured CSD as a function of gate voltages applied to $G_1$ and $G_2$. Peaks in the demodulated voltage $V_{\rm rf}$ correspond to Coulomb oscillations of the sensor dot. A red (blue) line cut highlights the  DQD's ICT transition with (without) the cascade process. Here we set $V_{G_3} = 355$ mV and $V_{G_4} = 100$ mV to keep the sensor dot in the multi-electron regime and to keep $D_4$ empty. 
(d) Simulated CSD diagram and charge occupancy of the triple dot tuned as in (c). For a full list of parameters used, see Appendix \ref{app:C_matrix}.}

\label{fig:Leti_device}
\end{figure}

We next extend rf electron-cascade readout to a different device and sensing architecture. Whereas the original implementation was sensed through a reservoir contact, here the cascade response is detected through the plunger gate of a sensor dot formed in a fully-depleted silicon-on-insulator (FDSOI) etched nanowire device, as shown in the simplified top view in Figure~\ref{fig:Leti_device}(a). Unlike in the previous experiment, the inductor used for reflectometry is not superconducting, and it is capacitively coupled to the plunger gate of the sensor dot (located underneath gate $G_{3}$) rather than galvanically connected to a reservoir. $G_3$ is wire-bonded to an 800-nH surface-mounted inductor for rf readout at the resonance frequency 191 MHz. Further details of the experimental setup can be found in Ref.~\cite{ansaloni2023gate}.

The device itself was fabricated in a FDSOI 300-mm-wafer process by LETI and implements a 2$\times$2 array of silicon QDs operating in the few-electron regime. 
In such a device, an undoped silicon channel is connected to highly doped source (S) and drain (D) reservoirs, with metallic polysilicon split gates partially overlapping the channel, under which QDs can form. 
An additional top gate is located 300 nm above the channel and can be used to change the overall tunneling rates within the array~\cite{ansaloni2020single}. 
The device is tuned in a triple QD configuration, with three dots under gates $G_{1,2,3}$ and no electrons below $G_4$. The multi-electron dot under $G_3$ is used as a sensor dot to measure the dot charge occupancy under $G_{1,2}$ in the few-electron regime. 
The fact that the transition is sensed from a gate rather than the reservoir can be accounted for by redefining $\alpha_{\rm casc} = \alpha_3 + \alpha_1 - \alpha_2$.

The measured CSD in Figure~\ref{fig:Leti_device}(c) is dominated by diagonal ridges of the demodulated rf signal $V_{\rm rf}$ that mostly correspond to the Coulomb peaks of the sensor dot, i.e. $D_3$ exchanging electrons with the source reservoir. Due to the capacitive coupling between $D_3$ and the other dots in the array, the ICTs are also visible, but much fainter. Of particular interest is the \textit{in-situ} ICT process and the \textit{rf electron cascade} process, highlighted by the blue and red lines, respectively. The red line cut crosses the transition where an electron is exchanged between $D_1$ and $D_2$, and simultaneously between the sensor dot and its reservoir (i.e. the electron cascade process). The blue line cut crosses the same ICT transition between $D_1$ and $D_2$, but without a change in the number of electrons in the sensor dot (i.e. no electron cascade process). Fig.~\ref{fig:Leti_device}(d) shows a simulation of the charge stability diagram in (c) with the electron occupation numbers below the gates $G_{1-4}$. The charge configurations at each side of the transition are calculated and confirm our hypothesis of the occurrence of an electron cascade event.

To quantify the signal amplification of the cascade process, we extract the SNR from the rf reflectometry signal along the red (blue) line cut from (c) with (without) the cascade process. The line cuts, interpolated using cubic splines, are plotted in Fig.~\ref{fig:Leti_device}(b). Applying the same methodology used in Sec.~\ref{sec:exp1} to extract the amplification factor yields an amplitude SNR enhancement of $ \sqrt{A}\approx 40$.

To summarize, in this section, we have shown in two different foundry-fabricated devices how, by tuning the sensor dot's chemical potential, the cascade process greatly amplifies the ICT signal.

\section{Conclusion and Outlook}

In this work, we have developed a general theoretical framework for \textit{rf electron-cascade} readout and extended its experimental demonstration to a second semiconductor QD architecture. Building on the original reservoir-sensed proof of principle, we have shown how an ICT in a target DQD can trigger a correlated DRT in an electrostatically coupled sensor dot, thereby converting the comparatively weak dispersive response of the ICT into the stronger ac response associated with a reservoir transition.

Exploiting the strong mutual capacitive coupling within a triple quantum-dot system, an ICT between two QDs can cause a third dot to exchange an electron 
with a reservoir. This correlated charge rearrangement bypasses energetically unfavourable intermediate charge states and enables a sensitive readout that leverages the larger alternating current associated with a reservoir transition. 

We derived analytic expressions for the cascade amplification factor $A$ in two limiting regimes. In the linear-response limit, 
$A \propto r^{-4}(\Delta/k_BT)^2$ for $\Delta \gg k_BT$, revealing a steep dependence on both the lever-arm ratio $r$ and the tunnel coupling $\Delta$. In the large-drive limit, the amplification approaches the geometry-dependent value $A=r^{-2}$ and becomes independent of temperature and tunnel coupling. These scaling laws identify three practical considerations for
maximising cascade amplification: minimising $r$ by increasing
the coupling to the sensed electrode, operating in the
linear-response regime, and reducing the electron temperature. The proposed mechanism is supported by measurements in two independent foundry-fabricated silicon devices, which exhibit lower-bound 
amplitude-SNR enhancements of $\sqrt{A} \approx 50$ and $\sqrt{A} \approx 40$, respectively.

More broadly, the cascade mechanism provides a route towards separating readout sensitivity from some of the constraints imposed by qubit control. Operations based on exchange interactions or Pauli spin blockade restrict the 
allowable interdot tunnel coupling within the target DQD. By introducing an additional sensor dot, the cascade architecture provides independent control parameters for optimising the readout response without directly modifying the operating conditions of the target DQD. This advantage comes at the cost of an additional device element and the need to maintain the sensor near the appropriate charge-degeneracy point. Future work should therefore investigate the robustness of the cascade against charge drift and finite-frequency effects, and directly test the predicted dependence on drive amplitude, temperature, tunnel coupling, and device electrostatics.

\section*{Acknowledgments}
This research was supported by the European Union’s Horizon 2020 research and innovation programme under grant agreements 951852 (QLSI), 101017733 (QuantERA II), 856526 (ERC grant), 101204890 (HORIZON-MSCA-2024-PF-01) under Horizon Europe, by the UK's Engineering and Physical Sciences Research Council (EPSRC) via the Cambridge NanoDTC (EP/L015978/1), the University College London DQT CDT (EP/S021582/1), and the Winton Programme for the Physics of Sustainability, and by the Novo Nordisk Foundation under Challenge Programme NNF20OC0060019 (SolidQ). M.F.G.Z. acknowledges a UKRI Future Leaders Fellowship [MR/V023284/1].

\section{Author Contributions}
T. M., L. P., G. A. O and M. F. G-Z. developed the theory of the cascade. T. M. and D.W developed the simulation tool (RF-Squad~\cite{murphy2025rfsquad}). T. M. performed the simulations. J. C-W. and R. C. C. L. acquired the data on device A. F. B., A. H., A. C. acquired the data on device B. F. K. and M.F. G-Z. supervised the work. All authors contributed to the writing of this manuscript.


\section{Competing interests}
J. C-W, R. C. C. L.
and M. F. G-Z. are inventors of a relevant patent application (WO2025093253A1). 

\appendix

\appendix

\section{Hubbard Model of a Triple Dot System}
\label{app:lorenzo}
To understand the \textit{rf electron cascade} phenomenon, we begin
with a spinless Hubbard model for the triple-quantum-dot (TQD) system,
described by the Hamiltonian
\begin{equation}
    H_{\mathrm{TQD}}
    =
    \sum_{j>i} U_{ij} n_i n_j
    +
    \frac{\Delta}{2}
    \left(
        d_1^\dagger d_2 + d_2^\dagger d_1
    \right)
    +
    \sum_{i=1}^{3} \varepsilon_i n_i ,
    \label{eq:H_hub}
\end{equation}
where $d_i^\dagger$ ($d_i$) is the fermionic creation
(annihilation) operator for dot $i$, and
$n_i=d_i^\dagger d_i$ is the corresponding occupation operator.
Here, $U_{ij}$ is the electrostatic interaction energy between dots
$i$ and $j$, and $\varepsilon_i$ is the on-site energy of dot $i$.
Coherent tunnelling is included only between dots 1 and 2, with
interdot hopping matrix element $t_c=\Delta/2$. Thus, $\Delta=2t_c$
denotes the full minimum energy splitting at the avoided crossing. $U_{ij}$ is the matrix representing the capacitive couplings within the QDs, and $\varepsilon_i$ their on-site energies \cite{RevModPhys.75.1}. Following the discussion in the main text, we assume that a coherent tunnel coupling $\Delta$ is present only between QDs 1 and 2 (those forming the DQD), while the interaction with the sensor ($D_3$) is purely electrostatic in nature. To include the presence of a charge reservoir (CR), we introduce the following Hamiltonians
\begin{align}
    H_{\rm CR} &= \sum_{\epsilon} \epsilon c_\epsilon^\dagger c_\epsilon\\
    H_{\rm CR-QD_3} &= \sum_{\epsilon} V_\epsilon c_\epsilon^\dagger d_3 + h.c.
    \label{eq:H_QD_R}
\end{align}
\noindent
where $c^\dagger$ ($c$) are the fermionic creation (annihilation) operators of the CR, and $V_\epsilon$ is the sensor QD-reservoir coupling, which, in the wide-band approximation, we consider independent of the reservoir energy $\epsilon$~\cite{Peri_2024,vonhorstig2024floquet}. We neglect the possibility of direct tunneling from the DQD to the CR.

To understand the evolution of the TQD system, we consider the reservoir to be a bath in thermal equilibrium (at temperature $T$), and derive a master equation via the Lindblad formalism of the form
\begin{align}
    &\mathcal{L} \rho = -i \left[H_{\rm TQD}, \rho\right] + \sum_p h \Gamma_p \mathcal{D}(L_p) \rho \label{eq:Lindbladian} \\
    &\mathcal{D}(L_p) \rho = L_p \rho L_p^\dagger - \frac{1}{2} \left\{L_p^\dagger L_p, \rho\right\}
    \label{eq:dissipator} ,
\end{align}
\noindent
where $\rho$ is the density matrix of the system, and $L_p$ take the name of \textit{jump} operators, describing (stochastic) decoherence processes occurring with rate $\Gamma_p$ \cite{manzano_short_2020}.
Moreover, we assume that the QD-CR coupling is weak enough that we can perform a rotating-wave (Born) approximation in the basis of eigenstates of Eq.~(\ref{eq:H_hub}) $\{\ket{\psi_l}\}$, whose eigen-energies are shown in Fig.~\ref{fig:Figure1}b.
In this approximation, it is possible to show that a coupling of the form in Eq.~(\ref{eq:H_QD_R}) gives rise to a series of jump operators of the form $L_{k,l}=\ket{\psi_k}\bra{\psi_l}$, with Fermi-golden-rule-like rates \cite{Peri_2024,Yamaguchi_Yuge_Ogawa_2017}
\begin{align}
    \tilde{\Gamma}_{\rm in, kl} &= \tilde{\Gamma}\left \vert \bra{\psi_k}d_3\ket{\psi_l}\right \vert^2 f(E_l-E_k - \mu)\\
    \tilde{\Gamma}_{\rm out, kl} &= \tilde{\Gamma} \left \vert \bra{\psi_k}d_3^{\dagger}\ket{\psi_l}\right \vert^2 (1- f(E_k-E_l - \mu))
\end{align}
\noindent
where $\mu$ is the chemical potential of the reservoir and $f$ its Fermi--Dirac distribution. Charge conservation mandates that $\tilde{\Gamma} = \sum_{k,l}\left( \tilde{\Gamma}_{\rm in, kl} + \tilde{\Gamma}_{\rm out, kl}\right)$ is the total tunnel rate, independent of energy.

Although the equations above are written as first-order transitions between eigenstates, this is equivalent to a cotunnelling process in which the transition does not reach the $(0,1,N)$ or $(1,0,N-1)$ configurations, since strong interdot electrostatic coupling makes both energetically inaccessible~\cite{Gustavsson2008Virtual}.

In the general case, this is the Lindblad equation governing the dynamics of the system, and gives rise to the full charge stability diagram of the TQD (e.g. Fig.~\ref{fig:Figure1}d) \cite{murphy2025rfsquad}.
However, to understand the electron cascade, we are mostly interested in the regime of large $D_2$--$D_3$ electrostatic coupling (i.e. $U_{23}\gg k_B T$). In this case, as amply discussed in the main text, all charge configurations are further in energy, and one may bias the system in such a way that the only the $\ket{\psi_{(0,1,N-1)}}$ and $\ket{\psi_{(1,0,N)}}$ eigenstates are relevant to the system's dynamics. Notably, these are not pure Fock states due to the finite coupling between $D_2$ and $D_3$.
If the electrostatic coupling is large enough, it is possible to ignore the thermal occupation of all other states, and it is possible to show that the dynamics of the system may be approximated as, in the language of Eq.~(\ref{eq:dP_main}),
\begin{equation}
    \frac{d}{dt} P(t) = - \Gamma P(t) + \Gamma_{\rm in} (t),
\end{equation}
\noindent
where $ \Gamma = \tilde{\Gamma} \left \vert \bra{\psi_{(0,1,0)}}d_3\ket{\psi_{(1,0,1)}}\right \vert^2$, and $\Gamma_{\rm in} (t) = \Gamma f(\varepsilon_3(t) - \mu)$.
This is the same equation governing a standard SEB \cite{Peri_2024}, from which the result in the main text and the above Appendix may be derived.

\section{Simulation parameters for charge stability diagrams}
\label{app:C_matrix}

\subsection{Simulation parameters for Fig.~2, ~4}

The charge stability diagrams shown in Figs.~2, ~4 were simulated using RF-Squad \cite{murphy2025rfsquad}, a constant-interaction model combined with an rf reflectometry solver.
Below we list all numerical parameters used in the simulations.

The dot--dot and dot--gate capacitance matrices (in units of aF) are

\begin{equation}
\mathbf{C_{dd}} =
\begin{pmatrix}
35.0 & -4.88 & -1.88 \\
-4.88 & 27.9 & -0.401 \\
-1.88 & -0.401 & 42.0
\end{pmatrix},
\end{equation}

\begin{equation}
\mathbf{C_{dg}} =
\begin{pmatrix}
0.897 & 0.145 & 0 \\
0.536 & 0.452 & 0 \\
0.0632 & 0.0198 & 2.01
\end{pmatrix}.
\end{equation}

In these capacitance matrices, the sensor dot corresponds to the third column, while the DQD correspond to the first two. The gate ranges used were

\begin{align}
G_1 &\in [0.10,\,0.35]~\mathrm{V},\\
G_2 &\in [0.15,\,0.40]~\mathrm{V},\\
G_3 &\in [0.30,\,0.32]~\mathrm{V}.
\end{align}

Each stability diagram was evaluated on a $200 \times 200$ grid in
$(G_1, G_2)$. The temperature was set to $T = 250~\mathrm{mK}$

\subsection{Simulation parameters for Fig.~5}

The dot--dot and dot--gate capacitance matrices (in units of aF) are

\begin{equation}
\mathbf{C_{dd}} =
\begin{pmatrix}
12.89 & -1.63 & -0.629 & -0.190 \\
-1.63 & 9.31 & -0.134 & -0.190 \\
-0.629 & -0.134 & 14.02 & -0.190 \\
-0.190 & -0.190 & -0.190 & 14.02
\end{pmatrix},
\end{equation}

\begin{equation}
\mathbf{C_{dg}} =
\begin{pmatrix}
0.898 & 0.146 & 0.0632 & 0 \\
0.537 & 0.453 & 0.0198 & 0 \\
0.0632 & 0.0258 & 2.016 & 0 \\
0.0632 & 0.0632 & 0.0632 & 2.016
\end{pmatrix}.
\end{equation}

The simulated gate values were

\begin{align}
G_1 &\in [0.001,\,0.14]~\mathrm{V},\\
G_2 &\in [0.00,\,0.20]~\mathrm{V},\\
G_3 &= 0.0350~\mathrm{V},\\
G_4 &= 0.015~\mathrm{V}.
\end{align}

The stability diagram was evaluated on a $200 \times 200$ grid in
$(G_1,G_2)$. The temperature was set to $T = 300~\mathrm{mK}$.

All CSD simulations, except when comparing the amplification factor, were normalized to values between zero and one.

\subsection{Simulation parameters for Fig.~3}
\label{app:fig3_parameters}

The amplification curves shown in Fig.~3 were calculated using the
quantum-capacitance model described in Sec.~II. The rf capacitances for the ICT and DRT were obtained by numerically evaluating Eq.~(4) of the main text.

\paragraph{Baseline parameters}

Unless otherwise stated, the simulations were performed with a reservoir
lever arm $\alpha_{\mathrm{R}} = 0.10$, lever-arm ratio
$r = 0.20$, temperature $T_0 = 50~\mathrm{mK}$, and tunnel coupling
$\Delta = 60~\mu\mathrm{eV}$.

\paragraph{Drive amplitudes}

Two drive amplitudes were used to represent the linear-response
and large-drive regimes. The linear-response drive amplitude,
\begin{equation}
    V_{\rm dev}=0.1~\mu\mathrm{V},
    \label{eq:Vdev_linear}
\end{equation}
was chosen such that
\begin{equation}
    \alpha_{\rm casc}eV_{\rm dev}\ll k_{\rm B}T,
    \qquad
    \alpha_{\rm in\text{-}situ}eV_{\rm dev}\ll\Delta.
    \label{eq:app_linear_drive}
\end{equation}
The large-drive amplitude,
\begin{equation}
    V_{\rm dev}=8~\mathrm{mV},
    \label{eq:Vdev_large}
\end{equation}
was chosen such that
\begin{equation}
    \alpha_{\rm casc}eV_{\rm dev}\gg k_{\rm B}T,
    \qquad
    \alpha_{\rm in\text{-}situ}eV_{\rm dev}\gg\Delta.
    \label{eq:app_large_drive}
\end{equation}
These inequalities are satisfied throughout the parameter ranges
used for the corresponding limiting-regime curves in Fig.~3.

\paragraph{Parameter sweeps}

For Fig.~3(a), the amplification was evaluated over the range
\begin{equation}
V_{\mathrm{dev}} \in [0.0,\,1.0]~\mathrm{mV}.
\end{equation}

For Fig.~3(b), the lever-arm ratio was varied logarithmically over
\begin{equation}
r \in [0.01,\,10].
\end{equation}

For Fig.~3(c), the temperature was swept in the range
\begin{equation}
T \in [30~\mathrm{mK},\,10~\mathrm{K}],
\end{equation}
and plotted as the dimensionless ratio $k_BT/\Delta$.

\section{Derivation of the amplification scaling laws}
\label{app:limits}

In this Appendix we derive the analytic limits of the amplification factor
$A = (C_{\textit{casc}}/C_{\textit{in-situ}})^2$ quoted in Sec.~II, and clarify the
crossover between the linear-response and large-drive regimes.

\subsection{Quantum-capacitance integral}
\label{app:Cq_integral}

The measured quantum capacitance is given as;
\begin{equation}
C_q
= \frac{e}{\pi V_{\rm dev}}
  \int_0^{2\pi} n_R\!\bigl(V_{\rm dev}\sin\theta\bigr)\sin\theta\;d\theta,
\label{eq:Cq_theta_app}
\end{equation}
where $\theta=\omega t$ and $t_{\rm int}$ spans an integer number of drive
periods.
The measured capacitance of either junction includes a lever-arm prefactor,
\begin{equation}
C = \alpha\, C_q,
\label{eq:C_alpha_app}
\end{equation}

For both the DRT and the ICT, the equilibrium occupation satisfies
$n_R(0)=\tfrac{1}{2}$, so it is convenient to decompose
\begin{equation}
n_R(v) = \tfrac{1}{2} + \tilde{n}_R(v),
\end{equation}
where $\tilde{n}_R$ is an odd function of $v$.
The constant $\tfrac{1}{2}$ integrates to zero against $\sin\theta$ in
Eq.~(\ref{eq:Cq_theta_app}), so $C_q$ is determined entirely by $\tilde{n}_R$.

\subsection{Linear-response regime ($eV_{\rm dev}\ll \text{k}_\text{B}T,\,\Delta/\alpha_{\rm in\text{-}situ}$)}
\label{app:linear}

When the drive amplitude is small, we expand to first order in $v$,
\begin{equation}
\tilde{n}_R(V_{\rm dev}\sin\theta)
\;\approx\;
n_R'(0)\,V_{\rm dev}\sin\theta.
\end{equation}
Substituting into Eq.~(\ref{eq:Cq_theta_app}) and using
$\int_0^{2\pi}\sin^2\theta\,d\theta=\pi$,
\begin{equation}
C_q \approx e\,n_R'(0).
\label{eq:Cq_linear_app}
\end{equation}
In the linear-response regime the quantum capacitance is therefore set by the
derivative of the occupation at $v=0$.

\subsubsection{Rf electron cascade}
\label{app:DRT_linear}

For the DRT, the occupation follows a Fermi--Dirac
distribution centered at the degeneracy point ($v=0$),
\begin{equation}
n_R^{\rm casc}(v)
= \frac{1}{\exp\!\bigl(\alpha_{\rm casc} e v / \text{k}_\text{B}T\bigr)+1}.
\label{eq:nR_DRT_app}
\end{equation}
Differentiating at $v=0$,
\begin{equation}
n_R^{\rm casc\,\prime}(0)
= -\frac{\alpha_{\rm casc}\,e}{4\text{k}_\text{B}T}.
\end{equation}
Including the lever-arm prefactor from Eq.~(\ref{eq:C_alpha_app}) and taking
the magnitude,
\begin{equation}
C_{\textit{casc}}^{\rm lin}
= \alpha_{\rm casc}\,e\,\bigl|n_R^{\rm casc\,\prime}(0)\bigr|
= \frac{\alpha_{\rm casc}^2\,e^2}{4\text{k}_\text{B}T}.
\label{eq:CDRT_lin_app}
\end{equation}
The two powers of $\alpha_{\rm casc}$ arise from one factor through the
chain rule on $\alpha_{\rm casc} ev$ in Eq.~(\ref{eq:nR_DRT_app}), and
one from the explicit prefactor in Eq.~(\ref{eq:C_alpha_app}).

\subsubsection{In-situ}
\label{app:ICT_linear}

For the ICT, the occupation is the thermal average over
the two-level avoided crossing,
\begin{equation}
n_R^{\textit{in-situ}}(v)
= \frac{1}{2}\!\left(1
  +\frac{\varepsilon_{\textit{in-situ}}(v)}{E(v)}
  \tanh\!\frac{E(v)}{2\text{k}_\text{B}T}\right),
\label{eq:nR_ICT_app}
\end{equation}
with $\varepsilon_{\textit{in-situ}}=\alpha_{\rm in\text{-}situ}\,e v$ and
$E=\sqrt{\varepsilon_{\textit{in-situ}}^2+\Delta^2}$.

At $v=0$, $E=\Delta$, and differentiating,
\begin{equation}
n_R^{\textit{in-situ}\,\prime}(0)
= \frac{\alpha_{\rm in\text{-}situ}\,e}{2\Delta}
  \tanh\!\frac{\Delta}{2\text{k}_\text{B}T}.
\end{equation}
Including the lever-arm prefactor,
\begin{equation}
C_{\textit{in-situ}}^{\rm lin}
= \frac{\alpha_{\rm in\text{-}situ}^2\,e^2}{2\Delta}
  \tanh\!\frac{\Delta}{2\text{k}_\text{B}T}.
\label{eq:CICT_lin_app}
\end{equation}

\subsubsection{Amplification in linear response}
\label{app:A_lin}

Forming the ratio from
Eqs.~(\ref{eq:CDRT_lin_app}) and (\ref{eq:CICT_lin_app}),
\begin{equation}
A^{\rm lin}
= \left(\frac{C_{\textit{casc}}^{\rm lin}}{C_{\textit{in-situ}}^{\rm lin}}\right)^2
= \frac{1}{r^4}
  \left(\frac{\Delta}{2k_BT\tanh(\Delta/2\text{k}_\text{B}T)}\right)^2,
\label{eq:Alin_app}
\end{equation}
where $r\equiv\alpha_{\rm in\text{-}situ}/\alpha_{\rm casc}$ is the lever-arm ratio.
This is Eq.~(\ref{eq:A_lin}) of the main text.

\paragraph{Limit $\Delta\gg \text{k}_\text{B}T$.}
Here $\tanh(\Delta/2k_BT)\to 1$ and
\begin{equation}
A^{\rm lin}\big|_{\Delta\gg \text{k}_\text{B}T}
\;\propto\;
\frac{1}{r^4}\left(\frac{\Delta}{\text{k}_\text{B}T}\right)^2.
\label{eq:Alin_hot_app}
\end{equation}
The relative power-response gain grows quadratically as the temperature is
reduced or the tunnel coupling is increased. These dependences have different
physical origins. Reducing $T$ increases the absolute cascade capacitance,
$C_{\rm casc}^{\rm lin}\propto 1/T$, whereas increasing $\Delta$ leaves
$C_{\rm casc}^{\rm lin}$ unchanged and instead suppresses the direct
\textit{in-situ} response, $C_{\rm in\text{-}situ}^{\rm lin}\propto 1/\Delta$.

\paragraph{Limit $\Delta\ll \text{k}_\text{B}T$.}
Here $\tanh(\Delta/2\text{k}_\text{B}T)\approx\Delta/(2\text{k}_\text{B}T)$, so
$\Delta/\tanh(\Delta/2\text{k}_\text{B}T)\to 2\text{k}_\text{B}T$, and
\begin{equation}
A^{\rm lin}\big|_{\Delta\ll k_BT}
\;\propto\;
\frac{1}{r^4}.
\label{eq:Alin_cold_app}
\end{equation}
The $\Delta$- and $T$-dependence cancel: the amplification saturates to a
temperature-independent floor set solely by the lever-arm ratio.

\subsection{Large-drive regime ($eV_{\rm dev}\gg\Delta/\alpha_{\rm in\text{-}situ},\,\text{k}_\text{B}T/\alpha_{\rm casc}$)}
\label{app:large}

When the drive amplitude is large enough to sweep both occupation functions
fully from $0$ to $1$ within each half-cycle, both $n_R^{\rm DRT}$ and
$n_R^{\rm ICT}$ approach step functions in opposite directions:
\begin{align}
n_R^{\rm DRT}(v)&\to\Theta(-v), \label{eq:nR_DRT_step_app}\\
n_R^{\rm ICT}(v)&\to\Theta(+v), \label{eq:nR_ICT_step_app}
\end{align}
where $\Theta$ is the Heaviside function.
Substituting either step function into Eq.~(\ref{eq:Cq_theta_app}) and
evaluating the integral analytically,
\begin{equation}
\int_{0}^{2\pi}
\Theta\!\left(\pm V_{\mathrm{dev}}\sin\theta\right)
\sin\theta\,d\theta
=
\pm 2.
\tag{C17}
\end{equation}
gives
\begin{equation}
C_q^{\rm DRT}\xrightarrow{V_{\rm dev}\to\infty}
-\frac{2e}{\pi V_{\rm dev}},
\qquad
C_q^{\rm ICT}\xrightarrow{V_{\rm dev}\to\infty}
+\frac{2e}{\pi V_{\rm dev}}.
\label{eq:Cq_step_app}
\end{equation}
Both quantum capacitances therefore share the same $1/V_{\rm dev}$ fall-off.
Including the lever-arm prefactors,
\begin{equation}
C_{\textit{casc}}^{\rm large}
= -\frac{2e\,\alpha_{\rm casc}}{\pi V_{\rm dev}},
\qquad
C_{\textit{in-situ}}^{\rm large}
= +\frac{2e\,\alpha_{\rm in\text{-}situ}}{\pi V_{\rm dev}}.
\label{eq:C_large_app}
\end{equation}

Forming the ratio of the two capacitances in Eq. \ref{eq:Cq_general} gives
\begin{equation}
A^{\rm large}
=
\left(
\frac{C_{\rm casc}^{\rm large}}{C_{\rm in\text{-}situ}^{\rm large}}
\right)^2
=
\left(
\frac{\alpha_{\rm casc}}{\alpha_{\rm in\text{-}situ}}
\right)^2
=
\frac{1}{r^2},
\label{eq:A_large_appendix}
\end{equation}
where $r=\alpha_{\rm in\text{-}situ}/\alpha_{\rm casc}$. Thus, in
the large-drive limit, the relative power-response gain is determined
solely by the lever-arm ratio and is independent of temperature,
tunnel coupling, and drive amplitude. This recovers Eq.~\eqref{eq:A_large} of the
main text.

\bibliography{bib2}
\end{document}